\documentclass[12pt]{iopart}
\usepackage{graphicx}
\begin{document}

\title[Absolute frequency measurements of $^{85}$Rb $n$F$_{7/2}$ Rydberg states]{Absolute frequency measurements of $\mathbf{^{85}}$Rb \textit{\textbf{n}}F$\mathbf{_{7/2}}$ Rydberg states using purely optical detection}

\author{L A M Johnson$^1$, H O Majeed$^1$, B Sanguinetti$^1$, Th Becker$^2$ and B T H Varcoe$^1$}

\address{$^1$ School of Physics and Astronomy, University of Leeds, Leeds, LS2 9JT, UK}
\address{$^2$ Max Planck Institute of Quantum Optics, Hans Kopfermann Str. 1 85748 Garching, Germany}
\ead{L.A.M.Johnson07@leeds.ac.uk}

\begin{abstract}

\noindent A three-step laser excitation scheme is used to make absolute frequency measurements of highly excited $n$F$_{7/2}$ Rydberg states in $^{85}$Rb for principal quantum numbers $n$=33-100. This work demonstrates the first absolute frequency measurements of rubidium Rydberg levels using a purely optical detection scheme. The Rydberg states are excited in a heated Rb vapour cell and Doppler free signals are detected via purely optical means. All of the frequency measurements are made using a wavemeter which is calibrated against a GPS disciplined self-referenced optical frequency comb. We find that the measured levels have a very high frequency stability, and are especially robust to electric fields. The apparatus has allowed measurements of the states to an accuracy of 8.0MHz. The new measurements are analysed by extracting the modified Rydberg-Ritz series parameters.      

\end{abstract} 


\newpage

\section{Introduction}\label{intro}

The accurate measurement of highly excited Rydberg level energies in the alkali atoms plays an important role in improving the accuracy of atomic models \cite{drake1991}. In most Rydberg spectroscopy experiments the atoms are detected via field ionization. However, in this study we use a method of purely optical detection in an ordinary vapour cell, which has been demonstrated in \cite{brandenberger2002,Mohapatra2008,thoumany2009,kubler10}. A vapour cell is a convenient and straightforward solution for finding Rydberg levels, that could potentially permit rapid advances in Rydberg spectroscopy. This technique presents a method of finding Rydberg states quickly, with a large signal to noise ratio and an apparent insensitivity to electric fields \cite{Mohapatra2008,thoumany2009}, which makes it particularly suited to studying high $\ell$ Rydberg states with large polarisabilities. It is therefore important to verify the ability to perform precision spectroscopy in such a setup. \\

\noindent Although there is a large body of work on precision interval and fine structure measurements of the different rubidium Rydberg series' \cite{harvey77,meschede87,li2003,han2006,afrousheh06}, measurements of the absolute energies of these levels are more difficult to carry out, and are therefore mainly limited to the lower $\ell$ states \cite{lee78,stoicheff79,Lorenzen83,sanguinetti2009}. It appears that absolute measurements of the $^{85}$Rb $n$F series have only been made once by Johansson in 1961 \cite{johansson61} for $n$= 4-8. However, as new tools are now available in laser spectroscopy, such as the optical frequency-comb technique, it is interesting to return to such measurements. In this work we wanted to demonstrate that precision laser spectroscopy measurements of Rydberg states could be effectively made using purely optical detection with a vapour cell sample.\\

\begin{figure}[h]
\begin{center}
\includegraphics[width=4.5cm]{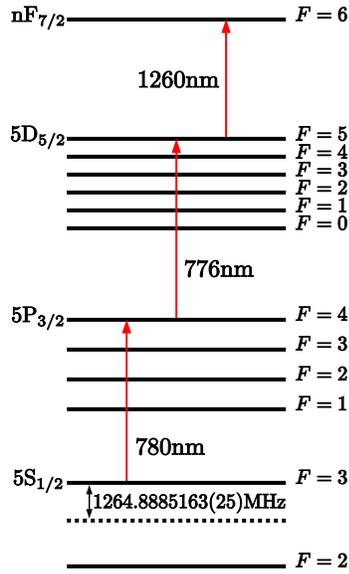}
\caption{The three step level scheme used to excite $^{85}$Rb $n$F$_{7/2}$ Rydberg states in this experiment.}
\label{level}
\end{center}
\end{figure}

\noindent During the experiment $n$F$_{7/2}$ Rydberg states between $n$=33-100 were excited in $^{85}$Rb using a three step laser excitation scheme identical to that outlined in \cite{thoumany2009,sanguinetti2009}. The three step level system, shown in figure \ref{level}, consists of a 780.24nm transition 5S$_{1/2}$ $F=3$ to 5P$_{3/2}$ $F=4$, a 775.98nm transition 5P$_{3/2}$ $F=4$ to 5D$_{5/2}$ $F=5$ and finally a 1260nm transition 5D$_{5/2}$ to $n$F$_{7/2}$.\\     

\noindent To observe excitations to Rydberg states, the first two step lasers are fixed at their respective transition frequencies and the absorption of the 780nm laser is monitored whilst the 1260nm laser is swept across the transition of interest. This technique involves the quantum amplification effect; due to the large differences in decay lifetimes of the three excited states of the system, the excitation of a single atom by the third step laser will hinder many absorption-emission cycles on the second step transition. This in turn will hinder a large amount of cycles on the strong first step cycling transition which can cause a measurable decrease in the first step absorption. Even for Rydberg atoms confined in a room temperature vapour cell, with the associated limitations of interaction time and interatomic collisions, this amplification factor can be large enough to observe significant changes in absorption \cite{thoumany2009}.\\

\noindent Optical pumping is applied on all three steps with $\sigma^{+}$ polarised light. Optical pumping on the first step transition ensures the second step laser only excites to the $m_{F}=5$ sublevel of the 5D$_{5/2}$ $F$=5 hyperfine state. Therefore the third step laser can only excite a single transition, the 5D$_{5/2}$ $F$=5 to $n$F$_{7/2}$ $F$=6. Having a well defined pathway to the Rydberg states is important because of the relatively small $\sim$10MHz hyperfine splitting of the 5D$_{5/2}$ level \cite{Nez1993}.

\section{Apparatus}\label{app}

\begin{figure}[h]
\begin{center}
\includegraphics[width=8cm]{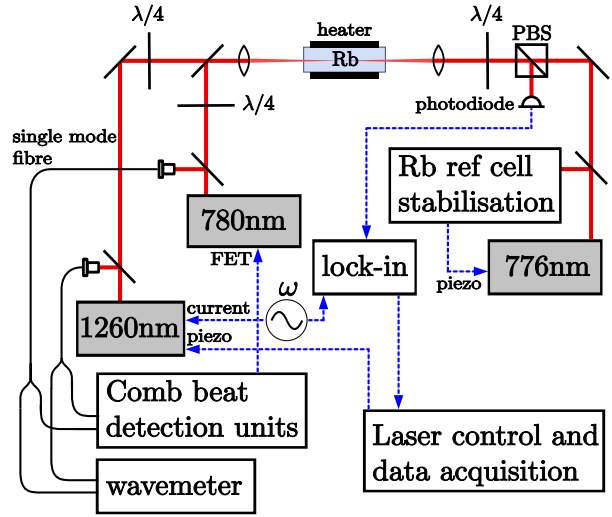}
\caption{The experimental setup used for measuring Rydberg state frequencies. The first step is phase locked to a self-referenced optical frequency comb and the second step is frequency locked using a separate rubidium reference cell. The first and third step laser light is transported to the comb and wavemeter using single mode optical fibres.}
\label{Setup}
\end{center}
\end{figure}

\noindent In the experimental setup, shown in figure \ref{Setup}, all three steps are excited using commercial tunable external cavity diode lasers, and associated electronics. The third step laser is broadly tunable across a range of 110nm using a precision stepper motor, this allows a large range of $n$F states to be accessed. The first and third step lasers are superimposed and co-propagate through a rubidium vapour cell of length 80mm. The second step laser travels through the same cell; it counter-propagates and overlaps with the first and third step lasers. Absorption of the first step laser is monitored using a conventional photodiode as the third step laser is swept across the 5D$_{5/2}$ to $n$F$_{7/2}$ transition of interest. Removal of the first step laser from the other two laser paths is carried out using a polarising beam splitter. The first two steps are circularly polarised using quarter wave plates, and the third step laser is circularly polarised using a broadband Fresnel rhomb. All three lasers are focused to a beam waist of $\sim$100$\mu$m inside the cell, which increases the available third step laser power density. The vapour cell is heated to a temperature of 60$^{\circ}$C to increase the atomic density in the cell and to therefore enhance the first step absorption. \\ 

\noindent In this experiment the first step laser Doppler selects those atoms which take part in the subsequent excitations, therefore it is important that the first step frequency is well known and well stabilised. Hence we stabilise this laser to a self-referenced frequency comb, by phase locking the beat note between the laser and a comb line to a stable direct digital synthesiser. The frequency comb repetition rate is adjusted such that the laser frequency is stabilised to 384\,229\,242.8 MHz, corresponding to the first step transition frequency from reference \cite{barwood1991}. All locking circuits are referenced to a GPS disciplined rubidium frequency standard. The comb system allows laser frequencies to be measured with an absolute accuracy of 10$^{-11}$. Fast feedback for the offset lock is supplied using a field-effect transistor connected to the laser diode. The stability of the first step lock was measured as less than 100Hz over all time scales relevant to this experiment. However, the absolute accuracy is limited to the measurement uncertainty  of 750kHz from Barwood \textit{et al} \cite{barwood1991}.\\ 

\noindent Before adding the third step laser to the system, we verified that efficient optical pumping was occurring on the first step transition by scanning the second step laser across the 5D$_{5/2}$ manifold, with the first step laser locked. The first step laser selects only zero velocity atoms, and therefore the second step laser scan showed a single and symmetric Doppler free peak in the first step absorption. This single peak, with a FWHM of 11.5MHz, corresponds to the reduced absorption of the first step laser as the second step laser excites the 5P$_{3/2}$ $F=4$ to 5D$_{5/2}$ $F=5$ transition. To confirm this we measured the absolute frequency of this transition using our frequency comb and added it to the first step locked frequency to get 770\,570\,284(1)MHz. This agrees with 770\,570\,284\,734(8)kHz from \cite{Nez1993}, obtained from two photon spectroscopy. This therefore demonstrates that the pathway to the Rydberg states is well understood. This scheme is also used to stabilise the second step laser with a separate room temperature vapour cell. By adding a small frequency modulation to the second step laser, and monitoring the first step absorption via a lock-in amplifier, an error signal is extracted. Using our frequency comb we verified that this second step frequency lock was repeatable to an absolute accuracy of 1MHz on a daily basis. \\

\begin{figure}[h]
\begin{center}
\includegraphics[width=7cm,angle=270]{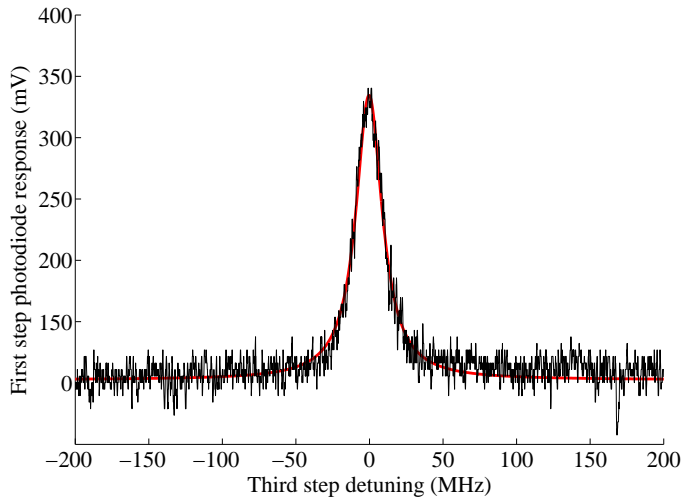}
\caption{A scan of the third step laser across the 33F$_{7/2}$ Rydberg state from an oscilloscope. The  vertical axis is the first step transmitted intensity from the photodiode. The frequency axis was calibrated with a Fabry-P\'{e}rot  resonator at 1268nm. The fitted curve is a Lorentzian with FWHM of 20MHz.}
\label{Lorentzian}
\end{center}
\end{figure}

\noindent We found that it is possible to detect lower $n$ states with a very good signal to noise ratio. Therefore to verify the line shape of the detected third step transitions the photodiode was monitored directly on an oscilloscope during a fast scan across the 5D$_{5/2}$ to 33F$_{7/2}$ transition. The trace is displayed in figure \ref{Lorentzian}. The scan was carried out in 10ms and the frequency axis was calibrated using a Fabry-P\'{e}rot resonator. The data fits a Lorentzian function with a linewidth of 20MHz.\\ 

\noindent To improve the detection sensitivity of third step transitions, a frequency modulation is added to the third step laser via the injection current, with a modulation amplitude of 15MHz and frequency of 6kHz. Detection of the first step absorption is carried out at the first harmonic using a lock-in amplifier with a time constant of 1 second. The free running third step laser is scanned by applying a linear voltage ramp to the laser Piezo using computer software and a Digital to Analogue converter interface. The free running laser stability was measured as less than 1MHz over one second, which is sufficient to carry out slow scans across the Rydberg transitions. As the third step laser is scanned, its absolute frequency is monitored using a WS7 High Finesse wavemeter. The wavemeter readings are recorded simultaneously using the same computer software.\\ 

\noindent We used our frequency comb to check the wavemeter's accuracy and stability across the range of third step laser wavelengths used in this experiment. We found that the wavemeter's stability stayed below 2MHz for times of $\sim$1000s. We also found that the wavemeter was able to maintain a day-to-day absolute accuracy of 6.2MHz across the 1254nm-1268nm range, when regularly calibrated at 780nm. Therefore, throughout this experiment the wavemeter is calibrated every 30 minutes to the comb-locked first step laser, to supply a direct frequency link with the comb.

\section{Results}\label{res}

\begin{figure}[h]
\begin{center}
\includegraphics[width=7cm,angle=270]{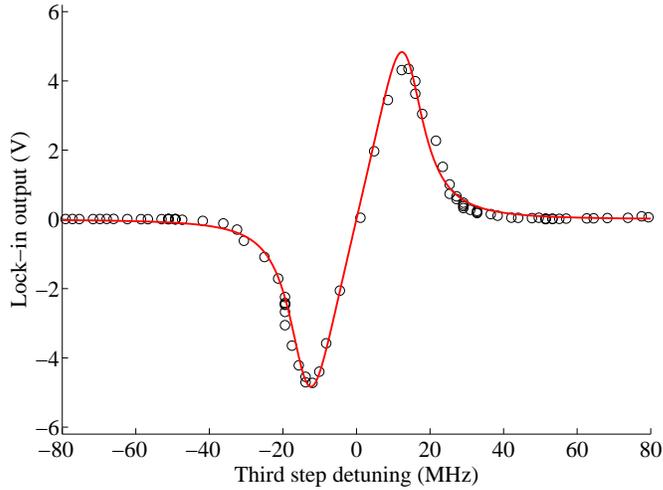}
\caption{A typical scan of the third step laser across the 33F$_{7/2}$ Rydberg state. The plot displays the demodulated first step absorption from the lock-in amplifier against the absolute frequency of the third step laser from the WS7 wavemeter.}
\label{Example}
\end{center}
\end{figure}

The third step transition absolute frequencies were collected for $n$=33-50 in intervals of one, and from $n$=50-100 in larger intervals of five. Fitting to the transition data was done using a Wahlquist first derivative function \cite{wahlquist1961}. The function is given by 

\begin{equation}
\label{wahlquist}
f(H_{\mathrm{\delta}}) = \frac{H_{\mathrm{\delta}}}{\vert H_{\mathrm{\delta}}\vert} \left(\frac{2}{H_{\mathrm{\omega}}}\right)^{2}  \frac{\sqrt{2 \gamma-u}} {2\sqrt{u-2}(u-\gamma)},
\end{equation}

\noindent where $\gamma= 1 + \beta^{2} + \alpha^{2}$, $u= \gamma + \sqrt{\gamma^{2} - 4 \alpha^{2}}$, $\alpha= H_{\mathrm{\delta}}/H_{\mathrm{\omega}}$ and $\beta = (\frac{1}{2} H_{1/2}/H_{\mathrm{\omega}})$. $H_{1/2}$, $H_{\mathrm{\omega}}$ and $H_{\mathrm{\delta}}$ are the FWHM, modulation amplitude and frequency detuning respectively. Figure \ref{Example} shows a typical scan across a Rydberg transition with the fitted profile from \eref{wahlquist}. We found that the linewidths of the detected third step transitions prevented resolving the $n$F$_{7/2}$ and $n$F$_{5/2}$ fine structure splitting in this experiment, which for $n$=33 to 100 is 4.35MHz to 0.16MHz respectively \cite{han2006}. However, the use of $\sigma^{+}$ light for the third step laser ensures only the $n$F$_{7/2}$ level is excited in this case.\\  
 
\noindent Ten traces were taken for each state in order to understand the repeatability of the measurements. It was found that on average the standard deviation of each set of ten scans was 2MHz with an accuracy limited by the short term drift of the wavemeter during the time taken to collect each set. The mean transition frequencies of the third step are summarised in the second column of table \ref{data}. The third column of this table displays the total 5S$_{1/2}$ to $n$F$_{7/2}$ frequency, measured from the center of mass of the 5S$_{1/2}$ ground states. These values were calculated by adding a constant value of 770\,571\,549.6MHz to the third step transition frequencies in column two, this frequency was computed from references \cite{Nez1993} and \cite{arimondo1977}.\\  

\begin{table}
\caption{\label{data}The third step transition frequencies $\nu_{3}$ and total 5S$_{1/2}$ to $n$F$_{7/2}$ transition frequencies $E_{n}$ for $n$=33-100. The total frequencies are measured from the center of mass of the two 5S$_{1/2}$ hyperfine ground states. The total accumulated error on all frequencies is 8.0MHz.}
\begin{indented}
\item[]\begin{tabular}{@{}llll}
\br
$n$      & $\nu_{3}$ & $ E_{n}$ \\
		 & (MHz)	 & (MHz)  	\\	
\mr

33 & 236\,429\,214 & 1007\,000\,764 \\
34 & 236\,604\,549 & 1007\,176\,099 \\
35 & 236\,765\,078 & 1007\,336\,627 \\
36 & 236\,912\,402 & 1007\,483\,952 \\
37 & 237\,047\,954 & 1007\,619\,503 \\
38 & 237\,172\,932 & 1007\,744\,481 \\
39 & 237\,288\,417 & 1007\,859\,967 \\
40 & 237\,395\,343 & 1007\,966\,892 \\
41 & 237\,494\,542 & 1008\,066\,092 \\
42 & 237\,586\,734 & 1008\,158\,283 \\
43 & 237\,672\,570 & 1008\,244\,119 \\
44 & 237\,752\,610 & 1008\,324\,159 \\
45 & 237\,827\,379 & 1008\,398\,929 \\
46 & 237\,897\,325 & 1008\,468\,875 \\
47 & 237\,962\,850 & 1008\,534\,399 \\
48 & 238\,024\,325 & 1008\,595\,874 \\
49 & 238\,082\,056 & 1008\,653\,605 \\
50 & 238\,136\,367 & 1008\,707\,917 \\
55 & 238\,364\,972 & 1008\,936\,522 \\
60 & 238\,538\,826 & 1009\,110\,376 \\
65 & 238\,674\,124 & 1009\,245\,673 \\
70 & 238\,781\,461 & 1009\,353\,011 \\
75 & 238\,868\,053 & 1009\,439\,602 \\
80 & 238\,938\,927 & 1009\,510\,477 \\
85 & 238\,997\,658 & 1009\,569\,208 \\
90 & 239\,046\,866 & 1009\,618\,416 \\
95 & 239\,088\,516 & 1009\,660\,066 \\
100 & 239\,124\,074 & 1009\,695\,624 \\

\br
\end{tabular}
\end{indented}
\end{table}

\noindent To study potential frequency offsets of the transitions caused by power shifts, pressure shifts or Zeeman shifts we took measurements of  both high and low $n$ states with a range of different first, second and third step laser powers, cell temperatures and opposite circular polarisations respectively. We also checked for errors from time delays in the data acquisition process by scanning the third step laser across the same transition in opposing directions. No repeatable shifts of the transition frequencies were found with increased laser powers or cell temperature and therefore potential offsets from these effects were not added as corrections but instead the spread of measurements were used to estimate a maximum error in each individual case. Neither Zeeman shifts nor time delay errors were detectable within the short term accuracy of the wavemeter and therefore these effects were assumed to give a negligible contribution to the uncertainty. The summarised error estimates are displayed in table \ref{errors}. The errors add in quadrature to give a total error of 8.0MHz.\\ 

\noindent Rydberg $n$F states are highly polarisable in external electric fields, with polarisabilities scaling as $n^7$ \cite{gallagher88}. To measure potential DC Stark shifts of the Rydberg states we applied electric fields of up to 30Vcm$^{-1}$ across the vapour cell and checked for frequency shifts of both the 33F$_{7/2}$ and 100F$_{7/2}$ transitions. In each case there was no measurable deviation. This unexpected observation was also made in references \cite{Mohapatra2008} and \cite{thoumany2009} when detecting Rydberg states in a cell. A screening of the Rydberg atoms inside the cell seems to be present, which makes them resilient to electric fields. This is a very positive effect as it allows precision spectroscopy of high $\ell$ states with no DC Stark shifts. 

\begin{table}
\caption{\label{errors}Estimated errors.}
\begin{indented}
\item[]\begin{tabular}{@{}llll}
\br
Source & Error \\
\mr
wavemeter calibration & 6.2MHz\\
first step frequency & 750kHz\\
second step frequency & 1.0MHz\\
pressure shifts & 2.7MHz\\
power shifts & 4.0MHz\\

\mr
TOTAL & 8.0MHz\\

\br
\end{tabular}
\end{indented}
\end{table}

\section{Analysis}\label{analysis}

Rydberg level energies are very well described by the Rydberg formula

\begin{equation}
\label{rydberg}
E_{n}= E_{\rmi} - \frac{R_{X}}{[n-\delta(n)]^{2}} = E_{\rmi} - \frac{R_{X}}{n^{*2}}  \ ,
\end{equation}

\noindent where $E_{\rmi}$ is the ionisation energy, $E_{n}$ is the excitation energy from the ground state to a state with principal quantum number $n$, $R_{X}$ is the Rydberg constant for the atom of interest, $\delta(n)$ is the quantum defect and $n^{*}$ is the effective quantum number. The quantum defect can also be written as a Ritz expansion

\begin{equation}
\label{ritz}
\delta(n) =  \delta_{0} + \delta_{2}t_{n} + \delta_{4}t_{n}^{2} + ... \ ,    
\end{equation}

\noindent where 

\begin{equation}
\label{ritzparam}
t_{n} =  \frac{1}{[n-\delta(n)]^{2}}  =  \frac{E_{\rmi}-E_{n}}{R_{X}} .
\end{equation} 

\noindent The data from this experiment was analysed using three different fitting  methods. The first two methods follow the same theme as \cite{martin79}, whilst the third method is a consistency check of the data with previous work. These methods are outlined in sections \ref{sec:meth1}, \ref{sec:meth2} and \ref{sec:meth3}. To aid in the analysis, five values of $E_{n}$ for $n$=4-8  were added to the data set from \cite{johansson61}. Weighted fitting was important to take account of the larger uncertainties on these older measurements. Throughout the analysis the Rydberg constant for rubidium 85 was taken as $R_{\mathrm{Rb}}$=10\,973\,660.672\,249$\times c$ from \cite{sanguinetti2009}.  

\subsection{Method 1}
\label{sec:meth1}

In method 1 the energy levels $E_{n}$ were fitted using a least squares fitting procedure to the formula:

\begin{equation}
\label{modified}
E_{n}= E_{\rmi} - \frac{R_{\mathrm{Rb}}}{ \left[ n- \delta_{0} - \delta_{2}t_{n} - \delta_{4}t_{n}^{2} - ...\right] ^{2}} \ .
\end{equation}

\noindent The fit algorithm balanced both sides of \eref{modified} to find the optimum parameters for $E_{\rmi}$, $\delta_{0}$, $\delta_{2}$, $\delta_{4}$,... The results from this fit are displayed in table \ref{fits} and the residuals are shown in figure \ref{Residuals}. Reference \cite{drake1991} describes in great detail how the series parameters extracted from this type of fit can explain physical properties of the Rydberg atom, such as the core polarisation. 

\subsection{Method 2}
\label{sec:meth2}

To remove the recursive nature of \eref{ritz} it is common to make the approximation 

\begin{equation}
\label{approx}
t_{n} \approx \frac{1}{(n-\delta_{0})^{2}} \ ,
\end{equation}

\noindent which when substituted into \eref{modified} gives a Rydberg-Ritz expression that can be evaluated with greater simplicity \cite{martin79}: 

\begin{equation}
\label{extended}
E_{n}= E_{\rmi} - \frac{R_{\mathrm{Rb}}}{[n-\delta_{0} - \frac{a}{(n-\delta_{0})^{2}} - \frac{b}{(n-\delta_{0})^{4}} -  ... ]^{2}} \ .  
\end{equation}

\begin{figure}[h]
\begin{center}
\includegraphics[width=7cm,angle=270]{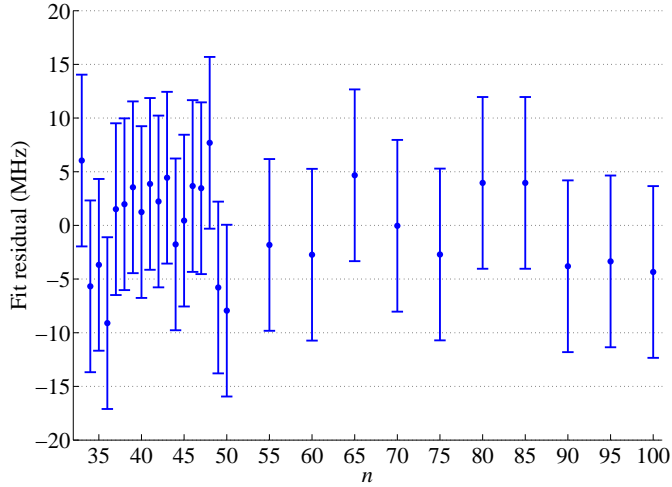}
\caption{The residuals for the $n$=33-100 states from the method 1 fitting routine. The error bars show the total accumulated error on each data point of 8.0MHz.}
\label{Residuals}
\end{center}
\end{figure}

\noindent The method 2 fit involved a direct least squares fit of \eref{extended} to the energy levels $E_{n}$. The results from this fit are displayed in table \ref{fits} where the $a$ and $b$ parameters are placed underneath the equivalent $\delta_{2}$ and $\delta_{4}$ parameters from the method 1 fit. It can be seen that the values of $E_{\rmi}$ and the series parameters extracted from the first two fitting methods agree to well within the uncertainties. The value of $E_{\rmi}$ from this work also lies within 2$\sigma$ of the previous value from \cite{sanguinetti2009}. An analysis of the residuals shown in figure \ref{Residuals}, from the method 1 fit, shows that the points are scattered around a mean of zero with a standard deviation of 4.4MHz. The states were measured across several days and therefore this spread comes mainly from the long term accuracy of the wavemeter.  \\

\noindent The Rydberg-Ritz formula in \eref{extended} has the significant advantage that it allows any energy level $E_{n}$ to be calculated with knowledge only of the principle quantum number $n$. In this manner \eref{extended} can be used with the relevant parameters in table \ref{fits} to predict the absolute energies of other rubidium $n$F$_{7/2}$ states outside the range of this experiment.

\subsection{Method 3}
\label{sec:meth3}

As a consistency check of this data we compared the Ritz series parameters extracted from our absolute measurements with those from the most recent relative interval measurements \cite{han2006}. For this fit we used an abridged version of \eref{extended}: 

\begin{equation}
\label{abridged}
E_{n}= E_{\rmi} - \frac{R_{\mathrm{Rb}}}{[n-\delta_{0} - \frac{a}{(n-\delta_{0})^{2}}]^{2}} \ .  
\end{equation}

\noindent This is the equivalent function which was used for fitting in reference \cite{han2006} and is an accurate approximation for $n \geq$20. For this reason we restricted this fit to the $n\geq$33 levels. The parameters from this fitting method are shown in table \ref{fits} with the values from \cite{han2006}.  The $a$ parameter is placed underneath the equivalent $\delta_{2}$ parameter from the method 1 fit.\\

\begin{table}
\caption{\label{fits}The fit parameters from the method 1, 2 and 3 fitting routines. Uncertainties are statistically derived from the fitting. The errors on the $E_{\rmi}$ values include possible systematic contributions outlined in table \ref{errors}. The equivalent parameters from reference \cite{han2006} are also shown for comparison with the method 3 results. }
\begin{indented}
\item[]\begin{tabular}{@{}lllll}
\br
					& $E_{\rmi}$ (MHz) 			&	$\delta_{0}$ 	& 	$\delta_{2}$    & 	$\delta_{4}$ \\
\mr
Method 1 			&	1010\,024\,719(8) 		& 	0.016\,473(14) 	& 	-0.0783(7)      & 	0.028(7)	\\
Method 2			&	1010\,024\,719(8)		&	0.016\,473(14)		&	-0.0784(7)  &   0.032(7)	\\
Method 3 & 1010\,024\,717(8) & 	 0.016\,40(8)		&	~0.00(9)& -		  \\
Reference \cite{han2006} & - &  0.016\,5437(7)     &	-0.086(7) & - \\

\br

\end{tabular}
\end{indented}
\end{table}

\noindent It can be seen that the $\delta_{0}$ and $a$ parameters from this fit agree at the 2$\sigma$ level with those from the previous work \cite{han2006}. Because our parameters are extracted from absolute measurements one does not expect as high an accuracy as from interval measurements, however absolute measurements do have the advantage that the ionisation energy $E_{\rmi}$ can also be extracted. The larger errors on the series parameters from this fit, as compared to the method 1 and 2 fits, arise because of the absence of lower $n$ states. This makes extracting higher order series parameters more difficult. For example, in \eref{extended}, for lower $n$ states the parameters $\delta_{0}$ and $a$ make a bigger contribution than for higher $n$ states, where $E_{n}$ becomes dominated by $E_{\rmi}$. As displayed in table \ref{fits}, the addition of the lower $n$ states from reference \cite{johansson61} greatly aided in the reliable extraction of the higher order parameters in the method 1 and 2 fitting routines. 

\section{Conclusion}\label{conc}

We have presented absolute frequency measurements of $n$F$_{7/2}$ Rydberg states in rubidium 85 to an accuracy of 8.0MHz. This is a factor 40 improvement over previous measurements of the $n$=4-7  $n$F$_{7/2}$ states \cite{johansson61} and gives measurements for a range of $n$F$_{7/2}$ states between $n$=33-100 for the first time. The Rydberg-Ritz series parameters which have been extracted from this work allow absolute energies of $n$F$_{7/2}$ states with higher or lower principle quantum number $n$ to be predicted with a comparable accuracy. Our new measurements also show consistency with results from recent microwave spectroscopy experiments \cite{han2006}. This work demonstrates that methods of Rydberg spectroscopy involving purely optical detection can be used very effectively to carry out precision measurements of Rydberg states in a simple way, and with extraordinary robustness to DC stark shifts. Not only is the set up simple to construct and maintain but it is easier to use than beam experiments, and Rydberg signals can be monitored in real-time on an oscilloscope. We believe that this experiment could be readily adapted to study other alkali metal atoms and could even be used to study such unusual features as Rydberg-Rydberg interactions and molecular states. In future work we hope to stabilise the third step laser to the transitions and directly count the laser frequency against a frequency comb. We estimate a potential 80$\times$ improvement in absolute accuracy can be made with this new approach. We also plan to study states with lower $n$ and $\ell$ by modifying the laser system. Carrying out these types of precision measurement on lower $n$ states would also allow quantum defects to be extracted with much greater accuracy. 

\section*{References}
\bibliographystyle{unsrt}
\bibliography{LukeJohnson}

\end{document}